\documentclass[3p,times,procedia]{elsarticle}

\usepackage{ecrc}

\volume{00}

\firstpage{1}

\journalname{Physics Procedia}

\runauth{}

\jid{phpro}

\jnltitlelogo{Physics Procedia}

\CopyrightLine{2011}{Published by Elsevier Ltd.}

\usepackage{amssymb}
\usepackage[figuresright]{rotating}

\begin{document}

\begin{frontmatter}

\dochead{}

\title{Measurements of the Time Structure of Hadronic Showers in a Scintillator-Tungsten HCAL}
\author{Frank Simon, for the CALICE Collaboration}

\address{Max-Planck-Institut f\"ur Physik, Munich, Germany\\
Excellence Cluster `Universe', Garching, Germany}

\begin{abstract}

For calorimeter applications requiring precise time stamping, the time structure of hadronic showers in the detector is a crucial issue. This applies in particular to detector concepts for CLIC, where a hadronic calorimeter with tungsten absorbers is being considered to achieve 
a high level of shower containment while satisfying strict space constraints. The high hadronic background from $\gamma\gamma \to {\rm hadrons}$ processes at 3 TeV in combination with the 2 GHz bunch crossing frequency at CLIC requires good time stamping in the detectors. To provide first measurements of the time structure in a highly granular scintillator-tungsten calorimeter, T3B, a dedicated timing experiment, was installed behind the last layer of the CALICE WHCAL prototype, a 30 layer tungsten scintillator calorimeter. T3B consists of 15 small scintillator cells with embedded silicon photomultipliers, read out with fast digitizers over a time window of 2.4 $\mu$s, and provides detailed measurements of the time structure of the signal. The offline data reconstruction performs an automatic gain calibration using noise events recorded between physics triggers and allows the determination of the arrival time of each photon at the photon sensor. The T3B setup, its calibration and data reconstruction, as well as first results of the time structure of the calorimeter response for 10 GeV pions recorded at the CERN PS confronted with Geant4 simulations using different physics lists are discussed. 

\end{abstract}





\end{frontmatter}

\section{Introduction}

For detector systems at CLIC \cite{Assmann:2000hg}, a hadronic calorimeter using tungsten absorbers is being considered to achieve a compact detector construction allowing the full containment of highly energetic jets while satisfying the space constraints imposed by the solenoidal magnet of the experiments. The high level of hadronic background from $\gamma\gamma \, \rightarrow \mathrm{hadrons}$ in combination with the bunch to bunch spacing of 0.5 ns, requires aggressive time stamping to limit the impact of this background on physics measurements.

 In the hadron calorimeter, the precision of timing is not only given by the detector technology and readout electronics, but also by the time structure of the hadronic cascade itself. This time structure in particular also defines the necessary integration times in the detectors, and thus the susceptibility to pile-up of background. The time structure of the shower is influenced by fast components of essentially instantaneous energy deposits from high-energy charged hadrons and from electromagnetic sub-showers, and by slow components, mainly due to neutrons and photons from nuclear processes.  Apart from the shower physics, the time structure of the calorimeter response is also influenced by the sensitive medium and by the detector electronics. The choice of the sensitive medium strongly affects the sensitivity of the detector to different components of the hadronic cascade, such as highly energetic charged particles, photons and neutrons, which each have a different time evolution. In addition, time constants of the medium itself, such as decay times in scintillators, affect the measurement.

Since the evaluation of the performance of calorimetry at a CLIC detector relies on simulations based on Geant4 \cite{Agostinelli:2002hh}, it is crucial to study how well the time structure of the detector response for hadronic showers in a tungsten calorimeter is reproduced by the simulations. To provide first measurements for a calorimeter using plastic scintillators, the Tungsten Timing Test Beam (T3B) setup has been installed in the CALICE scintillator tungsten calorimeter (WHCAL) prototype, which was taking data in a first test beam period at the CERN PS in Fall 2010. There, large data sets with pions at energies up to 10 GeV have been accumulated. In additional test beam campaigns in Summer 2011, this data set has been extended considerably at the CERN SPS, with energies all the way up to 300 GeV.

\section{The T3B Setup}

The T3B setup consists of fifteen $3\times3$ cm$^2$ scintillator tiles with a thickness of 5 mm, directly read out with 1 mm$^2$ Hamamatsu MPPC50P SiPMs with four hundred  $50\times50$ $\mu$m$^2$ pixels. The scintillator tiles have a ``dimple'' drilled into the side face at the SiPM coupling position to achieve a uniform response over the full active area. The layout of these scintillator tiles together with detailed response studies are presented in \cite{Simon:2010hf}. For T3B, the larger pixel MPPCs have been chosen because their increased photon detection efficiency  provides better sensitivity to small energies,  which is of particular importance for late energy deposits in the hadronic cascade. The T3B scintillator tiles provide a signal of approximately 27 photo electrons (p.e.) for minimum ionizing particles, including cross-talk and afterpulses of the photon sensor. 

\begin{figure}
\centering
\includegraphics[width=0.68\textwidth]{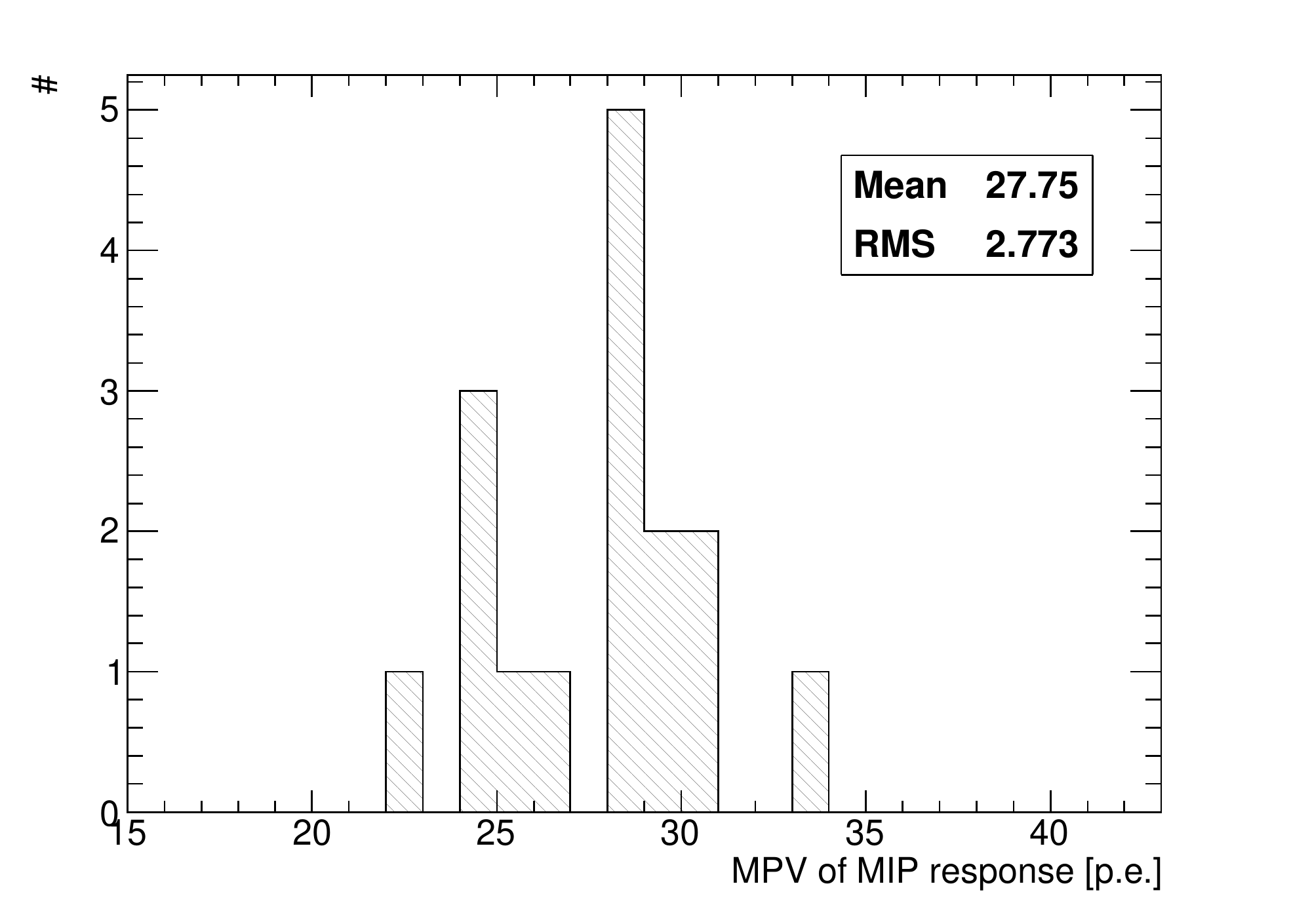}
\caption{Distribution of the most probable value of the response of the T3B scintillator tiles to penetrating electrons from a $^{90}$Sr source.}
\label{fig:MIPDistribution}
\end{figure}

The uniformity of the number of detected photons per minimum ionizing particle of all fifteen cells, measured on the test bench with a $^{90}$Sr source, is approximately 10\%. Figure \ref{fig:MIPDistribution} shows the distribution of the response to minimum ionizing particles at the nominal operation point for all cells used in the setup, plus one spare tile. This result demonstrates good reproducibility in the production of the scintillator cells for T3B. The photon sensors are read out with 4-channel USB oscilloscopes\footnote{PicoTech PicoScope 6403 (http://www.picotech.com/)} with 1.25 GS per second, using long acquisition windows of 2.4 $\mu$s per event to record the time structure of the energy deposits in the scintillator in detail.  Each SiPM is connected to a preamplifier board (as described in \cite{Simon:2010hf}), which then feeds the signal to the oscilloscope via coaxial cable. The bias voltage for each channel is adjusted by a resistor divider network, which is powered from one common high voltage source. The scintillator cells are individually packaged in light-tight tape. The preamplifier boards with packaged scintillator cells are mounted on a 2 mm thick aluminum plate and protected by a 1 mm thick aluminum top cover, forming a robust cassette.

\begin{figure}
\centering
\includegraphics[width=0.9\textwidth]{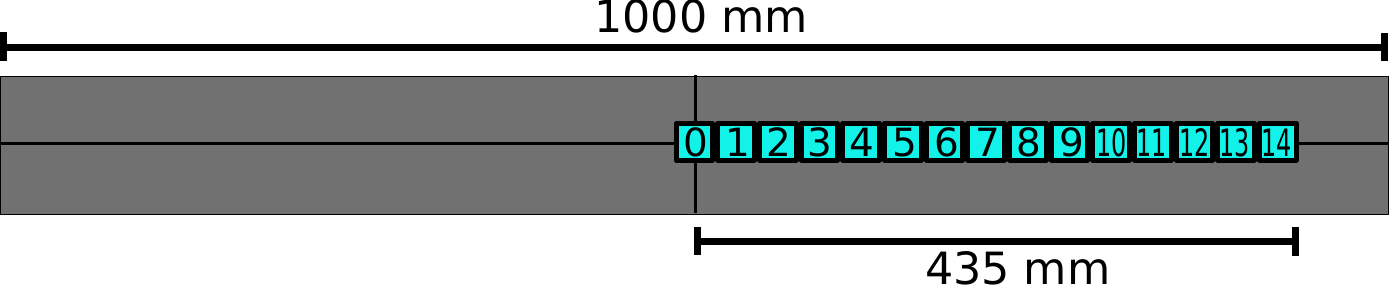}
\caption{Layout of the T3B scintillator tiles. From the nominal beam axis, the setup extends by 15 mm to one and 435 mm to the other side.}
\label{fig:T3BLayout}
\end{figure}

The T3B scintillator tiles are arranged in one row extending from the center of the calorimeter layer out to one side of the detector. The first tile is centered on the nominal beam position, thus the setup extends 15 mm beyond the nominal beam center on one and 435 mm on the other side, as shown in Figure \ref{fig:T3BLayout}. This permits the measurement of a full radial timing profile of the hadronic shower at the position of T3B, given sufficient statistics. The limited coverage however only allows averages over many events to be measured, and is not suitable for the study of the time evolution on an event by event basis.

In the 2010 test beam period, the T3B setup was installed behind the last layer of the CALICE WHCAL, which consists  of 30 tungsten absorber plates, 1 cm thick, supported by a 1 mm thick steel plate in addition. Behind each of the 30 absorber layers, one active layer of the CALICE analog HCAL \cite{Adloff:2010hb} is installed. After those 30 layers, additional slots without absorber plates are included in the mechanical structure, one of which is used for the T3B module, which is located at a depth of approximately 4 $\lambda_I$. 

During data taking, T3B is triggered by the main CALICE trigger, which will allow a matching of CALICE and T3B events in future analyses.For the readout, T3B uses four USB oscilloscopes totaling 16 channels, with the one channel not used for SiPM readout employed to monitor the signal from the trigger scintillators to allow the rejection of CALICE calibration triggers and the identification of double particle events. Between spills, T3B is taking dark rate events to allow a continuous monitoring of the SiPM gain. These are used for an automatic gain calibration during event reconstruction, discussed below. 

\section{Event Reconstruction}
\label{sec:Analysis}

For a first analysis of the 2010 dataset obtained at the CERN PS, only the highest energy data at 10 GeV have been considered, since this provides the highest activity in the rear of the calorimeter at the T3B position. This data sample contains approximately 718\,000 10 GeV $\pi^-$ events. For the analysis, T3B is used in standalone mode without attempting to correlate the events with CALICE WHCAL events to obtain additional information about the showers. The negative beam polarity is chosen to avoid a contamination with protons. At 10 GeV, the electron contamination of the beam is negligible. The muon contamination of the beam, which is not rejected, is around 10\%. The size of the trigger counters limits the influence of muons to the two central tiles of T3B. the size of the beam spot on the WHCAL front face is approximately 8 mm in both horizontal and vertical directions with an approximately Gaussian distribution, significantly smaller than the granularity of the T3B detector, allowing to use the cell index of the detector as a measure for the radial distance from the beam axis. 

\begin{figure}
\centering
\includegraphics[width=0.8\textwidth]{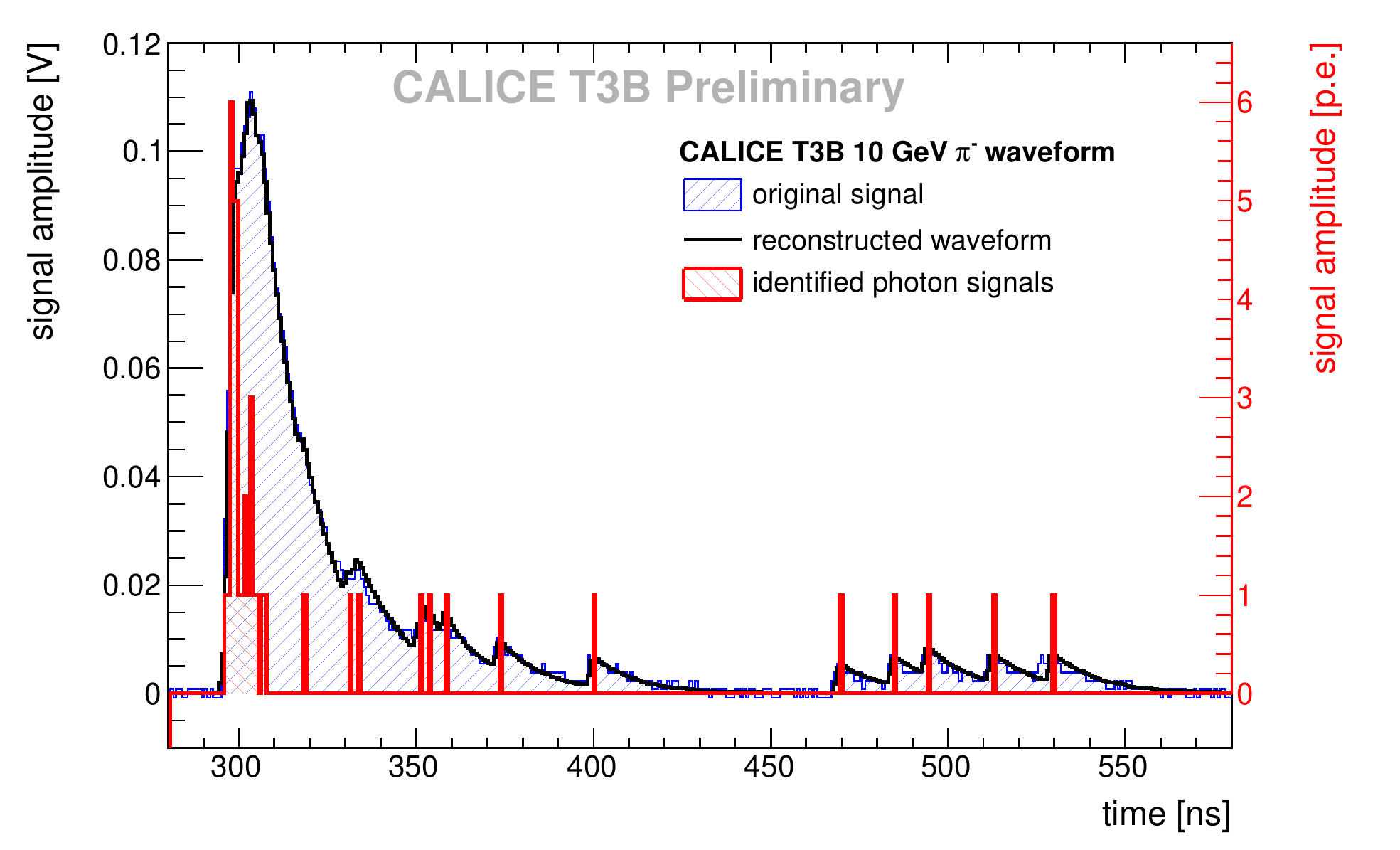}
\caption{Typical waveform with a high initial signal, decomposed into individual photon signals during the data analysis. Very good agreement of the original waveform and the reconstructed signal from standard single photo-electron distributions is observed. }
\label{fig:Waveform}
\end{figure}

The data are analyzed on a cell by cell level. As a first step, zero suppression based on pedestals determined on a spill-by-spill basis is applied. To reject events without significant signals, a threshold on the total zero-suppressed integral of the complete waveform equivalent to 0.3 MIP (corresponding to about 8 p.e.) is applied. 
To provide the best possible measurement of the time distribution of energy deposits in the scintillator, the goal of the event reconstruction is to determine the arrival time of each individual photon on the light sensor. This is done by consecutively subtracting single photon signals from local maxima detected in the waveform, until no maxima above approximately 0.5 p.e. remain. The single photon signals used for the subtraction are obtained from the calibration events taken between spills and are determined for each tile separately. To eliminate fluctuations from the vertical resolution of the oscilloscopes, these single photon waveforms are constructed by averaging many 1 p.e. dark noise pulses. Following this procedure, the final data analysis is then performed using the time distribution of reconstructed photons. 

This event reconstruction technique incorporates an implicit gain calibration, since possible cell-to-cell  or time-dependent gain differences lead to corresponding differences in the average single photon signals used in the analysis. The resulting number of photons is thus independent of the SiPM gain. This reference signal is refreshed every 10 spills, typically corresponding to time intervals of less than 5 minutes,  providing continuous automatic corrections of gain variations due to temperature changes. The temperature is monitored by one temperature sensor included in the T3B cassette, but that information is not used in the data reconstruction since the gain of each photon sensor is available. The temperature range of the data considered here is on the order of 0.8~$^{\circ}$C, resulting in gain variations of approximately 3\%. In addition to the gain, also the photon detection efficiency, and with that the number of detected photons per minimum-ionizing particle, depends on temperature. This dependence is not corrected for by the automatic calibration described here, and amounts to approximately 4\% over the temperature range of the data set at the operating conditions of the photon sensors in T3B. In the future, also this effect will be corrected for, using the measured gain of the photon sensors and the gain dependence of the signal for penetrating electrons from a $^{90}$Sr source measured on the test bench. 

Figure \ref{fig:Waveform} shows one example of a waveform decomposed into individual photons using the T3B reconstruction technique. To check the quality of this analysis, a waveform based on the identified photon signals is built up with the reference single photon signals and compared to the original waveform. The very good agreement between measurement and the reconstructed waveform demonstrates the quality of reconstruction. 

\begin{figure}
\centering
\includegraphics[width=0.99\textwidth]{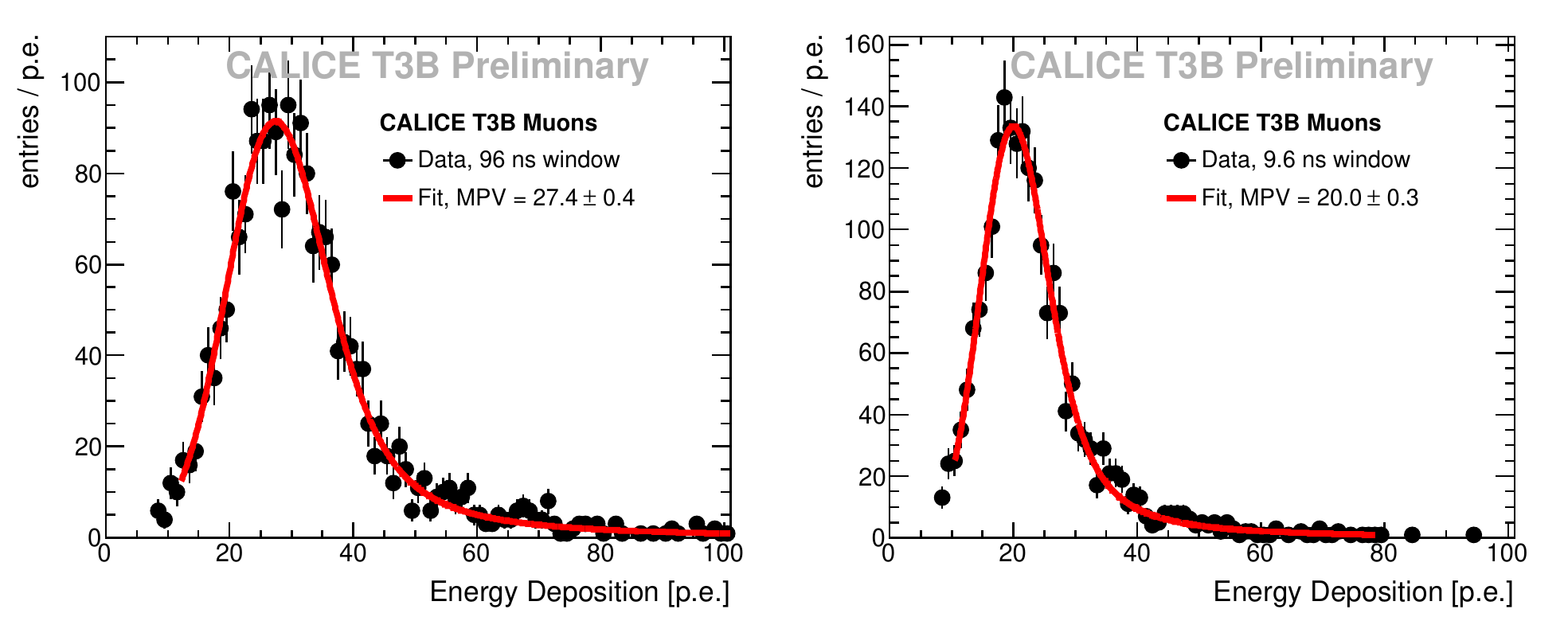}
\caption{Measured spectrum for muons in the central T3B scintillator tile, reconstructed by identifying the time of individual photon signals in the SiPM, for two different integration time windows: 96 ns from the first identified photon ({\it left}) and a time window of 9.6 ns ({\it right}). The distributions are fitted with a Landau function convolved with a Gaussian to extract the most probable value.}
\label{fig:Muons}
\end{figure}

Figure \ref{fig:Muons} shows the distribution of the energy reconstructed with this technique in the central tile of T3B for muons obtained in a special data run with an absorber in the beam line. The deposited energy is determined with two different integration windows, 96 ns from the first identified photons, shown in Figure \ref{fig:Muons} {\it left} and 9.6 ns, shown in Figure \ref{fig:Muons} {\it right}. The most probable value of both distributions are extracted by fitting a Landau function convolved with a Gaussian. The integration time has a considerable effect on the most probable value, which is reduced by almost 30\% from $27.4\,\pm\,0.4$ p.e. to $20.0\,\pm\,0.3$ p.e. for the reduced integration window. Also the width of the signal is reduced considerably. This is in part due to the reduction of contributions from afterpulses of the photon sensor, which account for approximately 2/3 of the observed difference, but may also include other effects such as scintillation of the reflective foil covering the tile, which may have longer time constants. The time resolution of the T3B system for single muons, including effects from the time jitter of the trigger system, is measured to be around 800 ps, demonstrating the possibility for sub-ns time resolution with this experiment.

\section{Analysis of Hadron Data}

In this first preliminary analysis, a simple, well defined quantity has been extracted: The time of first hit, given by the time when a cell saw its first energy deposit above a certain threshold. The analysis is performed on the 10 GeV $\pi^-$ sample, using the timing of each identified photon provided by the reconstruction procedure. Here, signals are considered where at least 8 p.e. are detected within 12 time bins (9.6 ns). This threshold corresponds to the equivalent of 0.4 MIPs. The time of hit is then taken from the timing of the second detected photon of that hit. It is observed that using the first instead of the second photon leads to additional jitter due to single p.e.~dark counts before the starting time of the real hit. At typical dark count rates of a few 100 kHz, the probability for such events within the integration time window is on the $10^{-3}$ level. The time of first hit for a T3B cell is given by the starting time of the first such hit in the waveform. 

\begin{figure}
\centering
\includegraphics[width=0.68\textwidth]{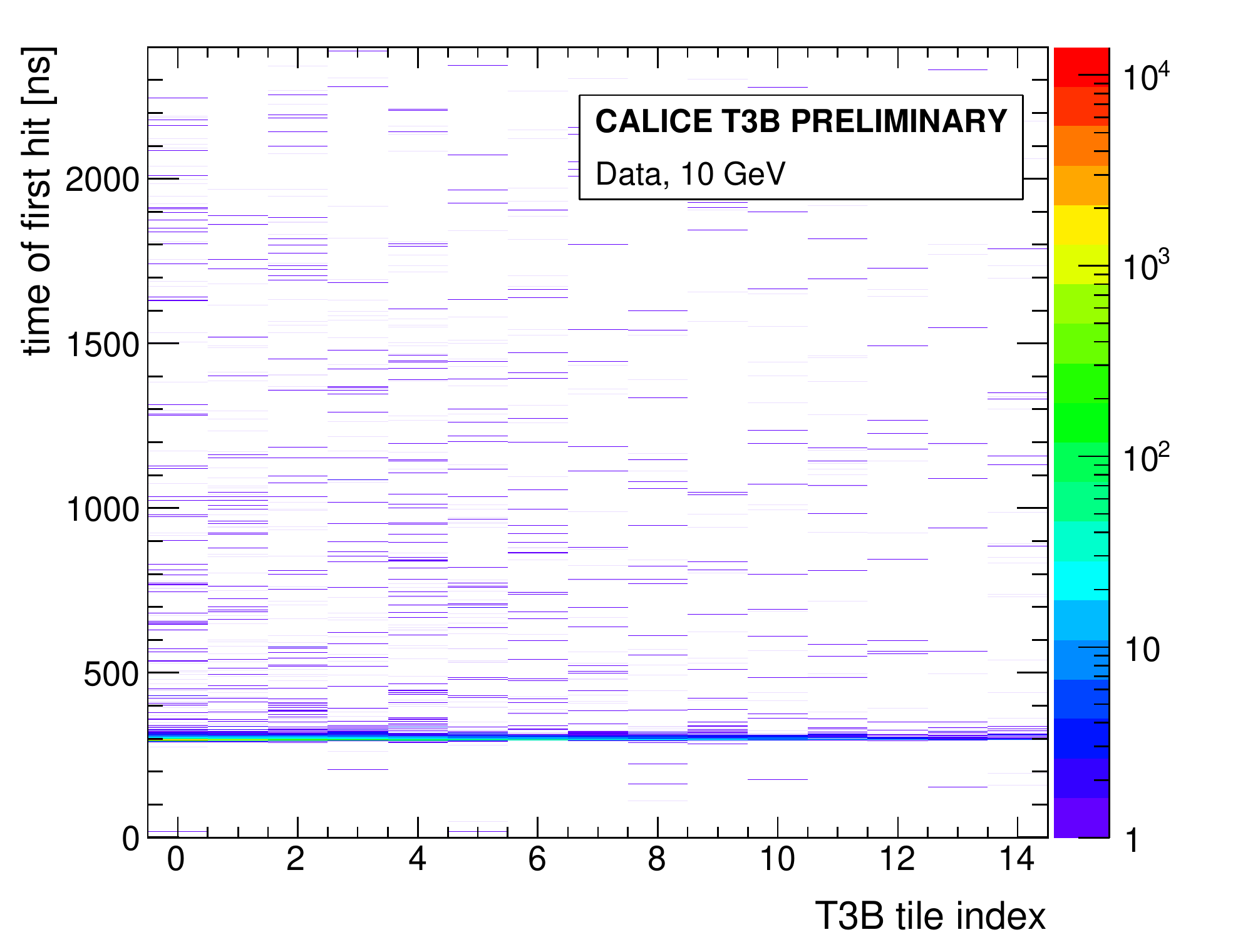}
\caption{Distribution of the time of first hit for all 15 T3B channels for 10 GeV $\pi^-$. The tile index corresponds to the radial position of the cell, with 0 on the beam axis, and 10 at a radial distance of 30 cm from the beam axis.}
\label{fig:Data2D}
\end{figure}

For the present analysis no cell-to-cell calibration is performed, leading to some uncertainty on the overall MIP scale for each cell. This uncertainty is on the 10\% level, as measured on the test bench, discussed above.  For the determination of the time of the first hit in a given cell, this missing calibration as well as changes of the signal amplitude due to temperature changes only contribute to higher order corrections from the application of thresholds, and do not influence the timing measurement itself. 

Figure \ref{fig:Data2D} shows the distribution of the time of first hit as a function of radial position, given by the T3B cell index, showing the expected cluster of events at early times in coincidence with the timing of the original beam particle, but also considerable late activity in the shower.

To confront the measurement with simulations, the geometry of the test beam setup has been implemented in Geant4 (version 4.9.3.p01). The reduction of the visible energy in the scintillator for slow, heavily ionizing particles is accounted for by using Birks' law \cite{Birks}. 800\,000 events, roughly corresponding to the available real data set, are simulated with the \texttt{QGSP\_BERT} and \texttt{QGSP\_BERT\_HP} physics lists \cite{Geant4Physics}. The latter provides additional high precision neutron tracking, and is expected to give an improved description of the shower evolution in heavy absorbers, while the former is the list mostly used in the simulation of LHC detectors and for linear collider optimization studies. For the events generated for the present study, the high precision neutron tracking increased the required CPU time for event simulation by approximately a factor of five. This CPU penalty can be significantly reduced in the future by turning off options in the high precision physics list accounting for neutron doppler shift,  which do not affect the results of the present study.

\begin{figure}
\centering
\includegraphics[width=0.99\textwidth]{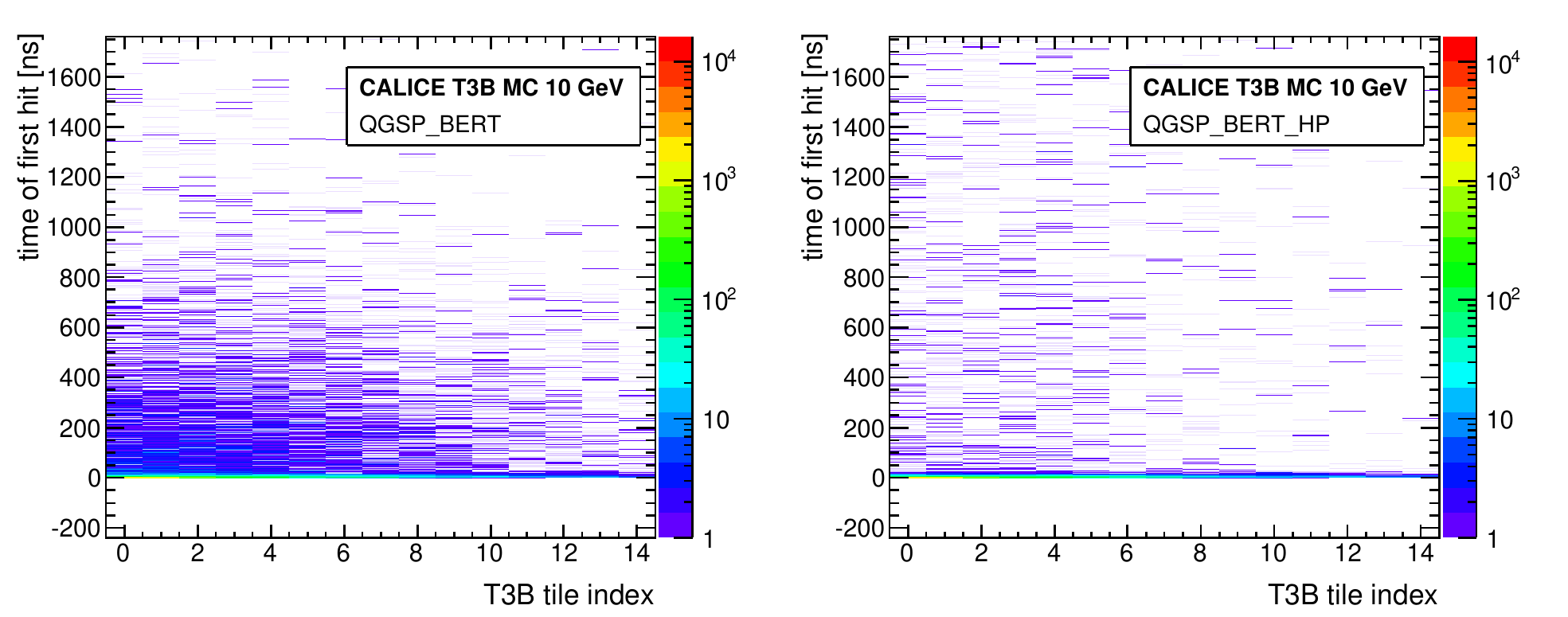}
\caption{Simulation of the distribution of the time of first hit for all 15 T3B channels for 10 GeV $\pi^-$ with \texttt{QGSP\_BERT} ({\it left}) and \texttt{QGSP\_BERT\_HP} ({\it right}). The tile index corresponds to the radial position of the cell, with 0 on the beam axis, and 10 at a radial distance of 30 cm from the beam axis.}
\label{fig:Sim2D}
\end{figure}

In the simulation, the deposited energy is converted into individual photons using the measured amplitude for and the known energy loss of minimum ionizing particles. The smearing in time originating from detector effects, such as scintillator time constants, photon travel times and the SiPM response, is taken into account by smearing the arrival time of the photons from the simulation by the measured time response of the system to instantaneous signals from muons. Figure \ref{fig:Sim2D} shows the simulated distribution of the time of first hit over all T3B tiles for the two considered physics lists, \texttt{QGSP\_BERT} and \texttt{QGSP\_BERT\_HP} for 10 GeV $\pi^-$.  These figures illustrate a striking difference in the late shower evolution for the two models. The delayed energy deposits are considerably reduced in the model including high precision neutron tracking.

\section{Results}

\begin{figure}
\centering
\includegraphics[width=0.68\textwidth]{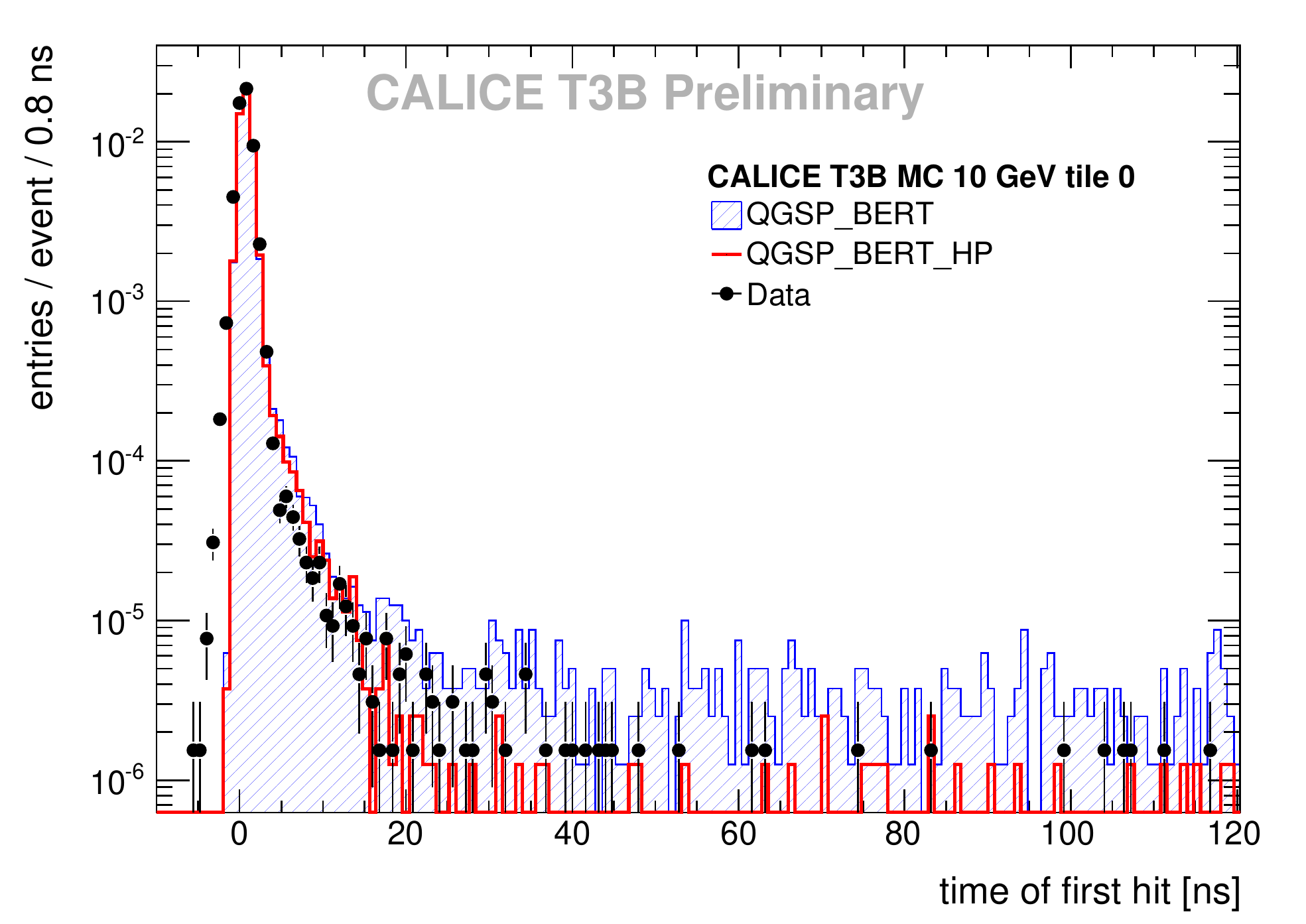}
\caption{Comparison of the time of first hit distribution in tile 0 simulated for 10 GeV $\pi^-$ with \texttt{QGSP\_BERT} and \texttt{QGSP\_BERT\_HP} with data.}
\label{fig:SimCompare}
\end{figure}

A direct comparison of the measured time of first hit with predictions by the two studied Geant4 physics lists is shown in Figure \ref{fig:SimCompare}. Here, the distribution in the central tile of T3B, which is close to the beam axis, is shown. While both physics lists reproduce the data very well in the main peak for instantaneous energy deposits in the first few ns, significant discrepancies are visible in the tail beyond 15 ns. Here, the hit distribution is considerably reduced in \texttt{QGSP\_BERT\_HP}  compared to \texttt{QGSP\_BERT}. While the former agrees well with the measurement, the latter significantly overestimates the late shower contributions. 

In order to provide a simple and robust comparison of these first results with simulations over the full radial extent of the T3B detector, the mean time of first hit is determined for each of the cells. This mean is formed by considering all measured hit times in a time window of 200 ns, starting 10 ns before the maximum of the distribution in T3B tile 0, and extending to 190 ns after the maximum. It thus included the full range shown in Figure \ref{fig:SimCompare} and extended by an additional 70 ns into the tails. This time window covers the time relevant for calorimetry at CLIC, where the duration for one bunch train is expected to be 156 ns, and is also comparable to the shaping time of 180 ns used in the front end electronics of the CALICE analog HCAL modules.

\begin{figure}
\centering
\includegraphics[width=0.68\textwidth]{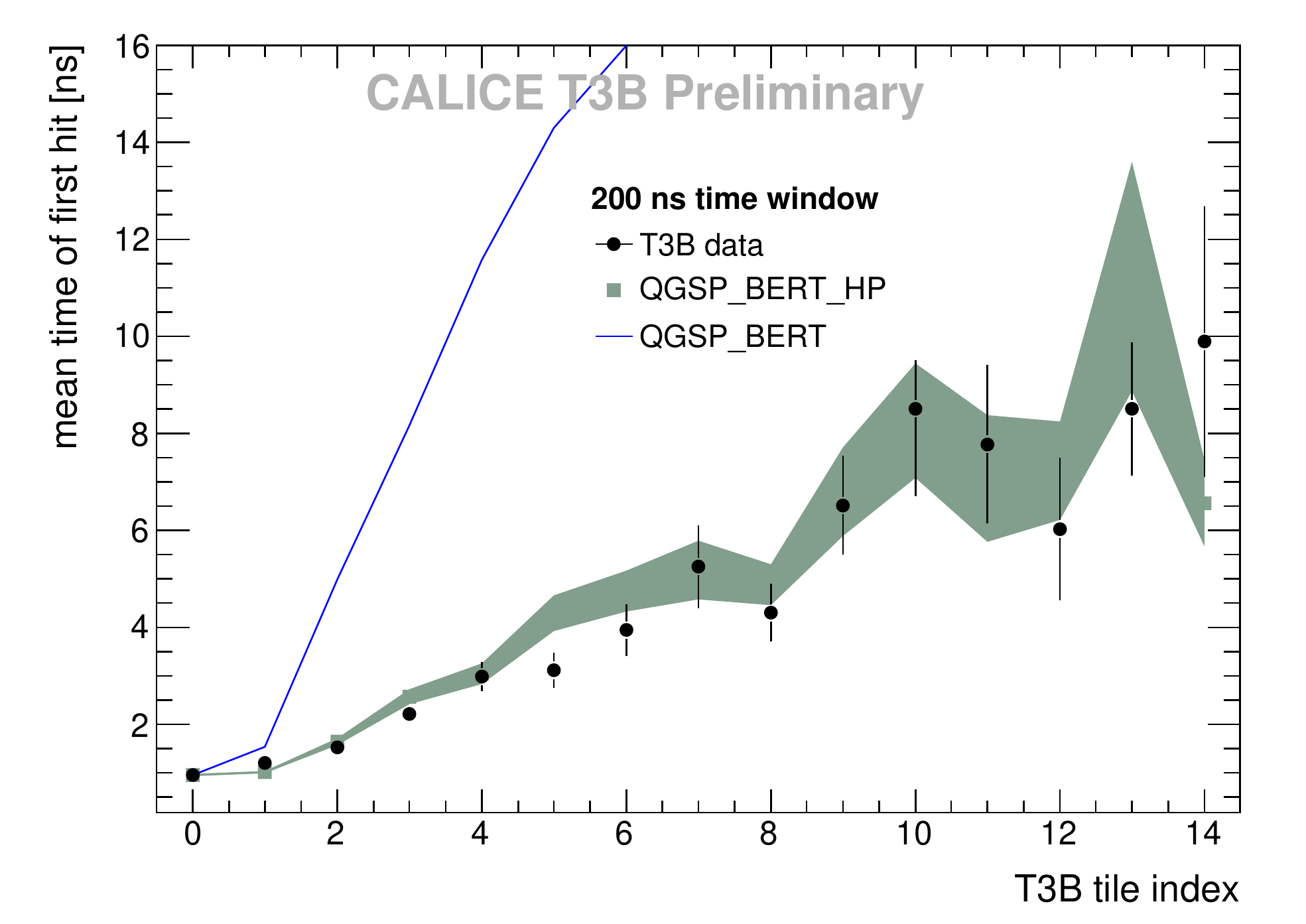}
\caption{Mean time of first hit for 10 GeV $\pi^-$ as a function of radial distance from the beam axis (a tile index of 10 corresponds to approximately 30 cm). The data are compared with simulations using \texttt{QGSP\_BERT} and \texttt{QGSP\_BERT\_HP}. The error bars and the width of the area in the case of  \texttt{QGSP\_BERT\_HP} simulations show the statistical error, while for  \texttt{QGSP\_BERT} the errors are omitted for clarity.}
\label{fig:MeanTimeResult}
\end{figure}

Figure \ref{fig:MeanTimeResult} shows the mean time of first hit as a function of the radial distance from the shower axis. The beam axis passes through T3B tile 0, so that a tile index of 10 corresponds to a distance of approximately 30 cm. The measurement is compared to the simulations with the two physics lists,  \texttt{QGSP\_BERT} and \texttt{QGSP\_BERT\_HP}.  While \texttt{QGSP\_BERT\_HP} gives an excellent description of the data, \texttt{QGSP\_BERT} shows very large discrepancies, with significantly overestimated late contributions at larger radii. This demonstrates the importance of the high precision neutron tracking in Geant4 for a realistic reproduction of the time evolution of hadronic showers in tungsten.

\section{Summary}

T3B, a dedicated experiment for first measurements of the time structure of hadronic showers in tungsten has been operated successfully for several test beam periods together with the CALICE analog scintillator-tungsten hadron calorimeter. T3B uses scintillator tiles directly read out with silicon photomultipliers, and is able to determine the time of arrival of individual photons at the photon sensor through specialized data reconstruction algorithms. First measurements of the time of first hit in 10 GeV $\pi^-$ showers, compared to Geant4 simulations, show very good agreement for the \texttt{QGSP\_BERT\_HP} physics list, while large discrepancies are observed for \texttt{QGSP\_BERT}. The use of high precision neutron tracking in the simulation reduces the number of late hits considerably, and brings the simulation in agreement with observation.

\bibliographystyle{elsarticle-num}
\bibliography{TungstenTiming}

\end{document}